\documentclass[fleqn,usenatbib]{mnras}

\usepackage{newtxtext,newtxmath}

\usepackage[T1]{fontenc}

\DeclareRobustCommand{\VAN}[3]{#2}
\let\VANthebibliography\thebibliography
\def\thebibliography{\DeclareRobustCommand{\VAN}[3]{##3}\VANthebibliography}

\usepackage{graphicx}	
\usepackage{amsmath}	
\usepackage{amssymb}	
\usepackage{listings}

\title[Three SB2 candidates]{Spectroscopic triples and a chance alignment. A solution for a problem of suspicious mass ratios for SB2s from Wilson method.}

\author[M. Kovalev et al.]{
Mikhail Kovalev,$^{1,2,3}$\thanks{E-mail: mikhail.kovalev@ynao.ac.cn}
Xuefei Chen$^{1,2,4}$
and Zhanwen Han$^{1,2,4}$
\\
$^{1}$Yunnan Observatories, China Academy of Sciences, Kunming 650216, China\\
$^{2}$Key Laboratory for the Structure and Evolution of Celestial Objects, Chinese Academy of Sciences, Kunming 650011, China\\
$^{3}$Sternberg Astronomical Institute, M. V. Lomonosov Moscow State University, Leninskie Gory, Moscow 119991, Russia\\
$^{4}$Center for Astronomical Mega-Science, Chinese Academy of Sciences, 20A Datun Road, Chaoyang District, Beijing 100012, China\\
}

\date{Accepted XXX. Received YYY; in original form ZZZ}

\pubyear{2023}

\def\kms{\,{\rm km}\,{\rm s}^{-1}}

\def\feh{\hbox{[Fe/H]}}

\newcommand{\teff}{T_{\rm eff}}
\newcommand{\rv}{{\rm RV}}
\def\Vmic{V_{\rm mic}}

\def\vsini{V \sin{i}}
\def\logg{\log{\rm (g)}}
\def\snr{\hbox{S/N}}
\newcommand{\ha}{\hbox{H$\alpha$}}

\begin{document}
\label{firstpage}
\pagerange{\pageref{firstpage}--\pageref{lastpage}}
\maketitle

\begin{abstract}
We selected three double-lined spectroscopic binary systems which have extreme mass ratios, if measured using the Wilson method. We analysed medium resolution spectroscopic observations and space-based photometry and find that all these systems are not SB2, but rather triple systems  and a chance alignment of another star with SB1 that have an unseen component. Therefore suspicious mass ratios determined by the Wilson method for some double-lined spectroscopic binary systems aren't correct as these systems are more complex.

\end{abstract}

\begin{keywords}
stars : fundamental parameters -- binaries : spectroscopic --  stars individual: J035146.75+252255.3, J061553.46+190014.8, J092306.86+431939.7
\end{keywords}



\section{Introduction}

Double-lined spectroscopic binaries are very useful astronomical objects since they allow us to estimate such important parameter as stellar mass. Mass ratio of the stellar components can be straight-forwardly measured for such systems using a linear fit of the line-of-sight velocities $\rv_{1,2}$, which is the so-called Wilson method \citep{wilson}. Unfortunately, this method sometimes produces extremely small or even nonphysical negative mass ratios \citep{kounkel}. 
\par
\begin{figure}
	\includegraphics[width=\columnwidth]{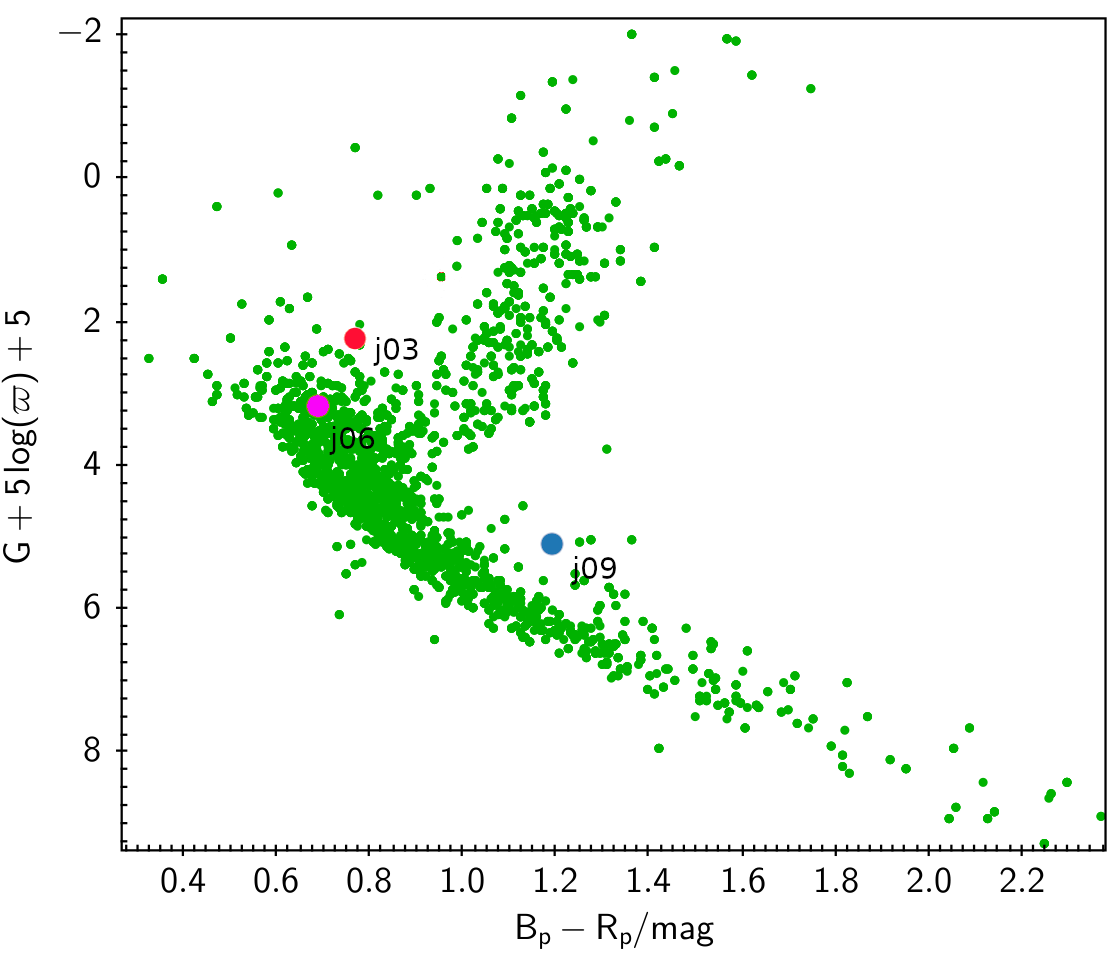}
    \caption{Hertzsprung-Russell diagram for all LAMOST MRS targets in the same field ($\sim4^\circ$) with j09 (blue circle) based on \protect\citet{gaia3} data. j03 and j06 are shown as red and magenta circles respectively.}
    \label{fig:hrd_gdr3}
\end{figure}

Many double-lined spectroscopic binary (SB2) candidates were identified in \cite{cat22} based on LAMOST (Large Sky Area Multi-Object fiber Spectroscopic Telescope) medium resolution spectra (MRS) \citep{mrs}. We arbitrary select three systems J035146.75+252255.3, J061553.46+190014.8 and J092306.86+431939.7 (hereafter j03, j06 and j09), where the Wilson method gives suspicious results, due to almost constant $\rv$ of the secondary spectral component. We list their literature info and designations in Table~\ref{tab:intro}. 
The first one - j03 is known eclipsing binary, located near the Pleiades, although it is not a cluster member \citep{j03membership}. It has a single-lined spectroscopic (SB1) orbital solution in Gaia DR3 \citep{gaia3} and clear double-lined spectrum in LAMOST-MRS. The second one - j06 is actually two stars separated by $2.5\arcsec$ which were observed by the same fibre of LAMOST-MRS\footnote{  fibre diameter is $\sim3.3\arcsec$} , so we can see the double-lined spectrum. The last one - j09 is the only system, where three spectroscopic components can be visually identified in the LAMOST-MRS spectrum. \cite{kounkel} found three spectral components in it's APOGEE DR16 spectrum \citep{apogee16}. Recently it was identified as a spotted star using ASSAS-SN photometry by \cite{asassn2023} and analysed for ellipsoidal variability in TESS photometry by \cite{green2023}. We show the position of all three systems on Hertzsprung-Russell diagram in Figure~\ref{fig:hrd_gdr3}, where they are all located higher than the main sequence for the field stars.

\begin{table*}
    \centering
    \begin{tabular}{c|ccc}
\hline
\hline
star & j03 & j06 & j09\\
\hline 
    LAMOST \protect\cite{cat22} & J035146.75+252255.3 & J061553.46+190014.8 & J092306.86+431939.7\\
    \hline
    Gaia DR3 \protect\cite{gaia3} & 66913361387671424 & 3373858954614854912  & 817543215158922368 \\
                                  & & 3373858958908468608 & \\
    $G$, mag & 11.56 & 12.77 & 12.90 \\
                                  & & 12.91& \\
    $\varpi$, mas & $ 1.3215 \pm0.0295$ &  $1.2137 \pm0.0175$& $2.8531 \pm0.0416$\\
                                        & &$1.1445 \pm0.0248$& \\
    SB1 orbit \\
    $P$, d & $5.73089\pm0.00032$ & & \\
    $t_0$, JD-2455197.5, d & $-2.51583\pm0.20445$ & & \\
    $e$ & $0.077\pm0.029$ & & \\
    $\gamma,\,\kms$ & $13.06\pm0.92$ & & \\
    $K,\,\kms$ & $44.46\pm1.29$ & & \\
    $\omega$, deg & $195\pm13$ & & \\
    \hline
    Variable Star indeX \protect\citet{varstarindex} &  HAT 260-0002375 (EBA)&  & ASASSN-V J092306.83+431939.7 (ROT)\\
    $P$, d & 5.73382176 &  & 1.3137\\
    Epoch HJD, d & 2457063.8446 & &  2458929.879\\
    \hline
    TESS input catalogue (TIC) \protect\cite{tic}&  84336450 &  718309429 &  99745836\\
                                  & & 718309430& \\
    ellipsoidal variability $P$, d \protect\cite{green2023} & & &$1.301848$\\

\hline
    \end{tabular}
    \caption{Literature data compilation for the selected three systems. For j06 there are two sets of data in Gaia DR3 and TESS. EBA - eclipsing binary of Algol type, ROT - spotted star.}
    \label{tab:intro}
\end{table*}

\par
In this paper we use available LAMOST-MRS spectra and additional photometrical data to explore these systems. After detailed analysis we conclude that they are actually not SB2s, but spectroscopic triples  j03, j09 and chance alignment of another star with j06 that have dark unseen component, possibly a white dwarf. 
The paper is organised as follows: in Section~\ref{sec:obs} we describe the observations. Section~\ref{sec:methods} presents our analysis and results. In Section~\ref{discus} we discuss the results. In Section~\ref{concl} we summarise the paper and draw conclusions.

\section{Observations}
\label{sec:obs}
\subsection{Spectra}

LAMOST (also known as Guo Shou Jing telescope) is a 4-meter quasi-meridian reflective Schmidt telescope with 4000 fibres installed on its 5-degree-FoV focal plane. These configurations allow it to observe spectra for at most 4000 celestial objects simultaneously (\cite{2012RAA....12.1197C, 2012RAA....12..723Z}).
 All available spectra for our three targets were downloaded from \url{www.lamost.org}.	We use the spectra taken at a resolving power of R$=\lambda/ \Delta \lambda \sim 7\,500$. Each spectrum is divided on two arms: blue from 4950\,\AA~to 5350\,\AA~and red from 6300\,\AA~to 6800\,\AA. We convert the  heliocentric wavelength scale in the observed spectra from vacuum to air using \texttt{PyAstronomy} \citep{pya}. 
For the short period ($P<2$ ) systems j06 and j09, we analysed spectra taken during individual 20 minutes exposures, but for j03 we used spectra stacked for the whole night. In total we have 13, 60 and 63 spectra for j03, j06 and j09 respectively, where the average signal-to-noise ratio ($\snr$) of a spectrum ranges from 10 to 72 ${\rm pix}^{-1}$ for individual exposures and from 69 to for 240 ${\rm pix}^{-1}$ for stacked spectra.

\subsection{Photometry}

The Transiting Exoplanet Survey Satellite \citep[TESS][]{tess} mission observed all three systems, but for j03 there is high-quality light curve (LC) from Kepler K2 \citep{polarK2}, which is available on the MAST portal\footnote{\url{https://mast.stsci.edu/}}, so we use it due to much higher precision. This LC covers time interval BJD=2457061:2457132 d. 
For j06 LCs are not available on the MAST portal yet, therefore we use \texttt{eleanor} \citep{eleanor,astrocut} to extract the LC datasets from sectors 43, 44 and 45, covering the time interval BJD=2459474:2459551 d. We use default settings and slightly clip the edges of the LC, as they have some processing artifacts. 
For j09 we use TESS LC from sector 21 reduced by Quick Look Pipeline (QLP) \citep{tess1,tess2} and available on the MAST portal. This LC covers the time interval BJD=2458870:2458896 d.
 
\section{Methods and results} 
\label{sec:methods}

\subsection{Spectral analysis}
Our spectroscopic analysis includes two independent stages: 
\begin{enumerate}
    \item analysis of individual observations by binary and single-star spectral models, where we normalise the spectra and make a rough estimation of the spectral parameters, see brief description in Section~\ref{sec:ind}. 
    \item spectral disentangling of the stacked spectra using {\sc FD3} code \citep{fd3norm,fd3}, which is described in Section~\ref{sec:sepa}.
\end{enumerate}

\subsubsection{Individual spectra.}
\label{sec:ind}

A detailed description of the spectral analysis is given in \citet{cat22}. In short, we used synthetic spectral models from \url{nlte.mpia.de}, generated to be good representation of the single-star LAMOST-MRS spectra, see Appendix~\ref{sec:payne}.
The normalised binary model spectrum is generated as a sum of the two Doppler-shifted normalised single-star model spectra ${f}_{\lambda,i}$ scaled according to the difference in luminosity, which is a function of the $\teff$ and stellar size. We assume both components to be spherical.

 The observed spectrum is compared with model and normalised using using \texttt{scipy.optimise.curve\_fit} function. Each spectrum is fitted independently, unlike \cite{tyc,j0647}, as we are unable to put constraints on the mass ratio during the simultaneous multiple epochs fitting. 
\par
\begin{figure*}
	\includegraphics[width=0.9\textwidth]{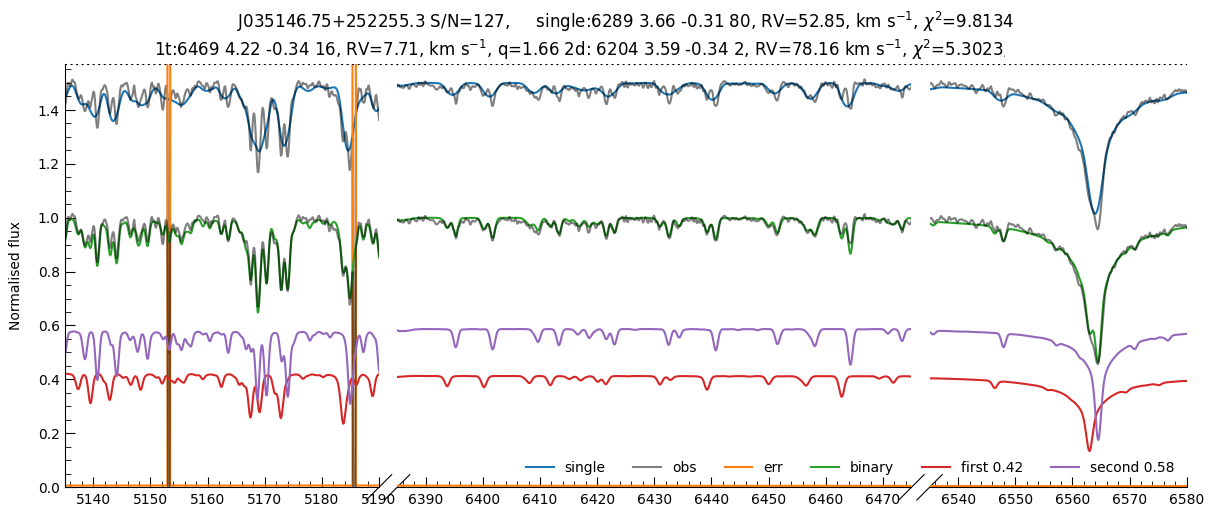}
    \includegraphics[width=0.85\textwidth]{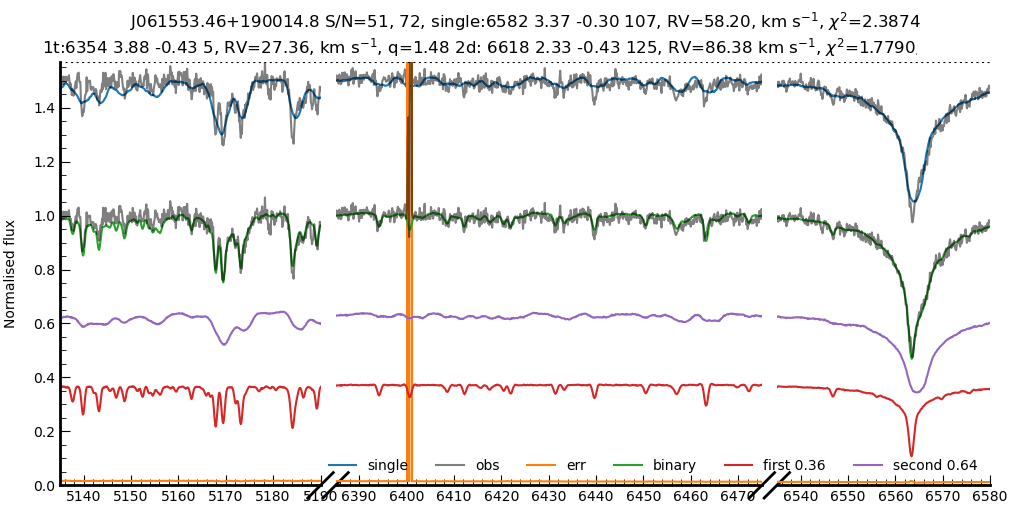}
    \includegraphics[width=0.85\textwidth]{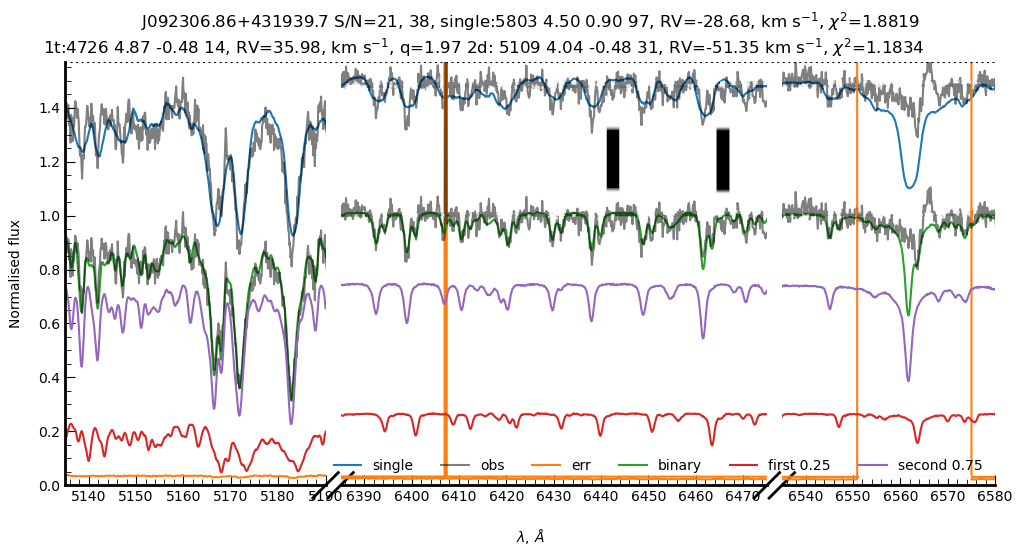}
    \caption{Example of the binary model (green line) and single-star model (blue line + 0.5 offset) fitting for LAMOST-MRS spectra for j03 (top), j06 (middle) and j09 (bottom). The observed spectrum and it's error are shown as a gray and orange lines respectively. The primary component is shown as the purple line, the secondary as a red line. The third component is visible only for j09 as an emission line at $\ha$ and two weak lines at $\lambda=6442,\,6465$~\AA (black blocks). Spectral parameters ($\teff,~\logg,~\feh,~\vsini$) and $\rv$ from single star model fit and binary model fit are shown in the titles.} 
    \label{fig:spexampl}
\end{figure*}

 
We show representative examples of our spectral fits for all three systems in Figure~\ref{fig:spexampl}. As can seen in all plots, the binary model is able to fit the observed spectrum much better than a single-star model, although for j09 a small contribution from a third component is also visible. Also note that for j09 we exclude the spectral region around the $\ha$ line as it has emission associated with the third component.   

\subsubsection{Spectral disentangling}
\label{sec:sepa}

\begin{figure*}
	\includegraphics[width=0.9\textwidth]{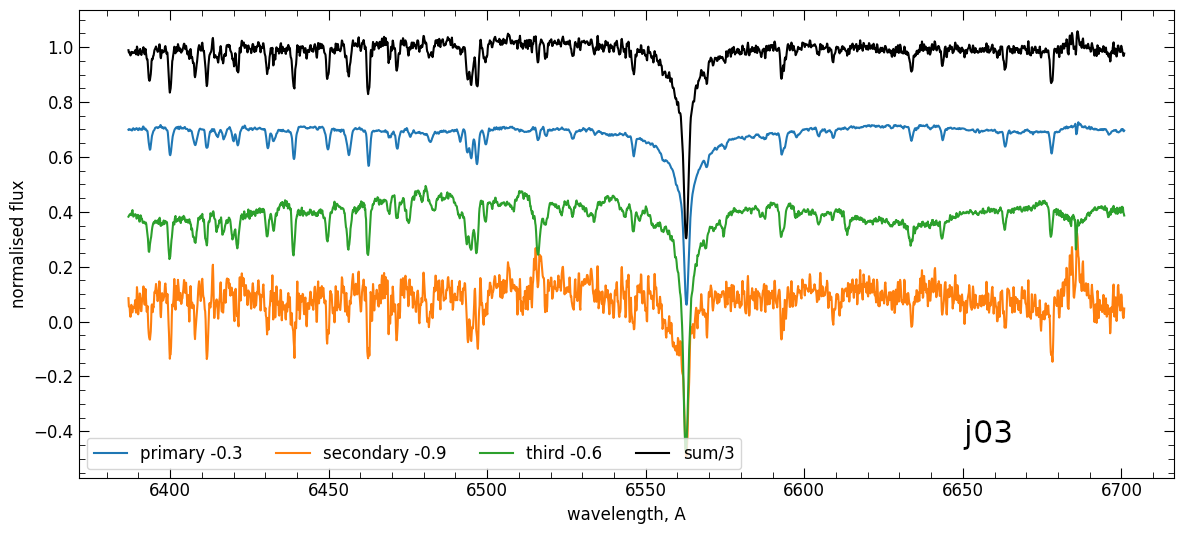}
    \includegraphics[width=0.9\textwidth]{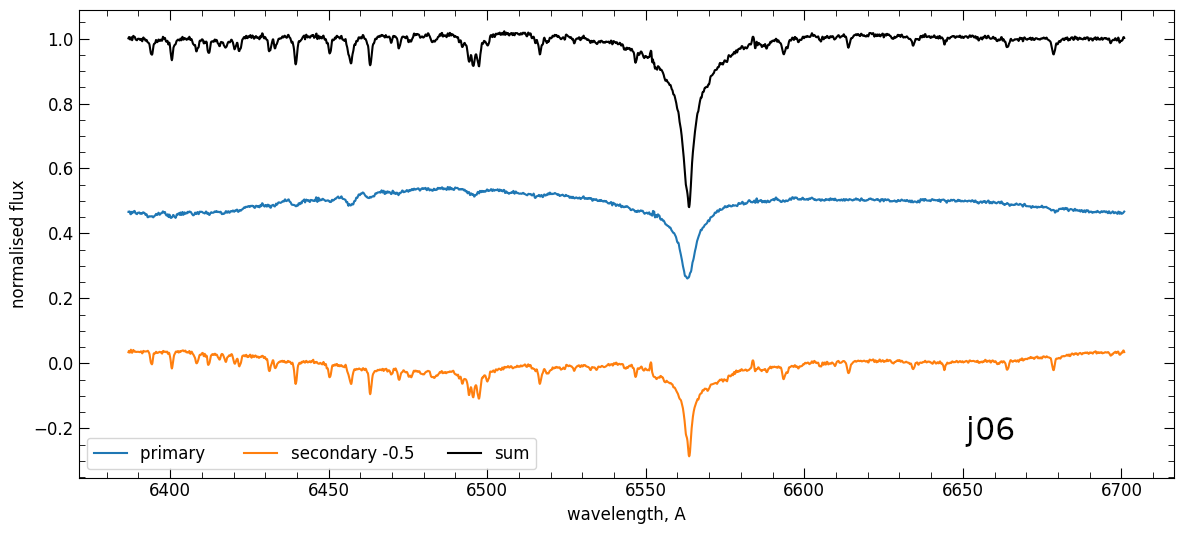}
    \includegraphics[width=0.9\textwidth]{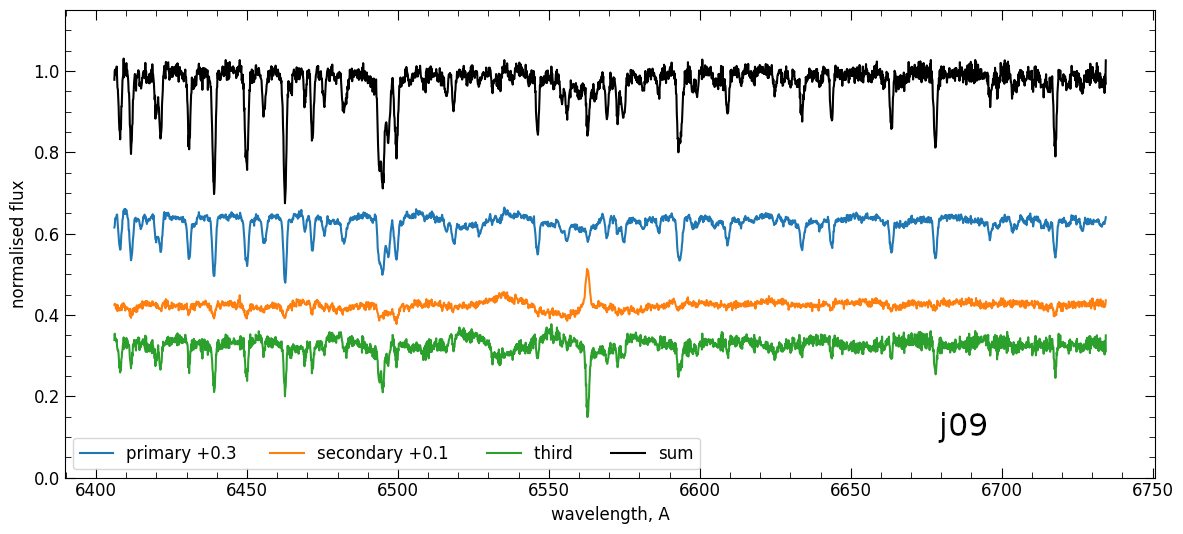}
    \caption{Example of the spectral separation for j03 (top), j06 (middle) and j09 (bottom) . The total spectrum is shown as a black line. The primary component is shown as the blue line, the secondary as an orange line. The third component is shown as a green line.}
    \label{fig:sepfd3}
\end{figure*}

Similarly to \cite{j0647} we use the Fourier spectral disentangling code {\sc FD3} by \cite{fd3} for all three systems. This method requires clean spectra, without artefacts (i.e spikes from cosmic rays), thus we use it with stacked spectra. It can separate up to three spectral components and find the Keplerian orbit for an inner close binary. We use the blue and red spectral arms only for j03, while for j06 and j09 only the red arm is used. At first we try to separate the three spectral components for j06, but find no good solution, thus this system was analysed with only two components. For j03 and j09 we successfully find three components. For j06 and j09 we use light factors $LF=1$ for all spectra, but for j03, thanks to the LC solution by {\sc JKTEBOP} (see Section~\ref{sec:lc}), we set $LFs$ according to the light contribution of each component in the spectra. Therefore only for j03 spectral components are normalised to unity, while for j06 and j09 components the baseline is 1/2 and 1/3 respectively.      
We show the resulting spectral components in Figure~\ref{fig:sepfd3} and list orbital solution's parameters in Table~\ref{tab:fd3}. As seen in the case of j03 all components have relatively narrow spectral lines, with the secondary component having deeper lines, but it's spectrum is too noisy as it contributes $\sim 5$ per cent of total light. In j06 the primary component has spectral lines with significant rotational broadening, while lines of the other component are relatively narrow. For j09 we can see emission in the $\ha$ line of the secondary, while this line is almost absent in the primary. Also note that spectral lines in the primary are more broadened than in the third component, which possibly rotates much slower. For all these plots we tried our best effort to remove smooth, kind-of-sinusoidal undulations in the spectral components, taking into account useful advise from Sa{\v s}a Iliji{\'c} (private communication) and \cite{fd3norm}.

\begin{table}
\begin{center}
\caption{{\sc FD3} orbital solutions for stacked spectra in the red arm. For j03 we also provide results for the blue arm spectrum. Unfortunately {\sc FD3} doesn't provide any uncertainties. }
\begin{tabular}{lccc}
\hline
Parameter & j03 & j06 & j09\\
\hline
$P$, d & 5.73128 & 0.47656 & 1.31059\\
(blue) & 5.73157 & & \\
$t_0$, HJD d & $2458806.995$ &$2458904.049$ & $2458904.353$\\
(blue) & $2458806.378$ & & \\
$K_1,\,\kms$ & 62.96 & 44.63 & 83.94\\
(blue) & 62.97 & & \\
$K_2,\,\kms$ & 98.78 & & 114.27\\
(blue) & 100.69 & & \\
$e$      &    0.003 &   0.006 & 0.004\\
(blue) & 0.001 & & \\
$\omega\, \degr$ & 134.5 & 96.5 & 70.0\\
(blue) & 96.3 & & \\

\hline
\label{tab:fd3}
\end{tabular}
\end{center}
\end{table}

\subsection{Orbital fitting}
To get the orbital solution we select 13, 60 and 63 RVs of the primary components in j03, j06 and j09 respectively. We collect all RV measurements in Table~\ref{tab:rvs} and use them
 to fit circular orbits with the generalised Lomb-Scargle periodogram (\texttt{GLS}) code by \cite{gls} :
\begin{align}
    {\rm RV}_A(t)=\gamma- K_A \sin \left (\frac{2\pi}{P}(t -t_0) \right ),
\end{align}
where $\gamma$ - is the systemic velocity, $P$ - is the period, $t_0$ - is the conjunction time, $K$ - is the radial velocity semiamplitude.
 We also fit to a Keplerian orbit and find that eccentricity is nearly equal to zero for all three systems, so a circular orbit is a valid assumption.
\par

\begin{figure}
    \centering
    \includegraphics[width=\columnwidth]{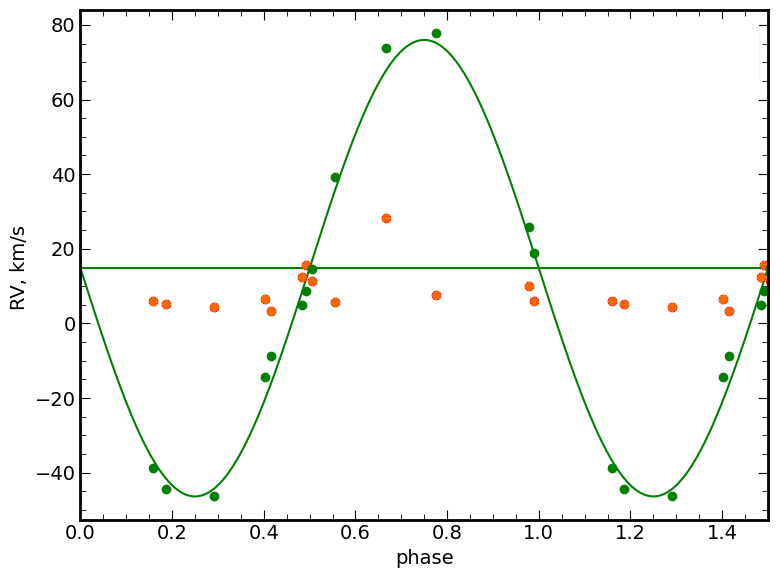}
    \caption{  Circular orbit fit (green line) for the primary component ($\rv_1$) of j03 (green circles). Note that the secondary spectral component ($\rv_2$) (orange circles) has blue shift relative to the systemic velocity.}
    \label{fig:rv03}
\end{figure}
The circular orbit fit for j03 is shown in Figure~\ref{fig:rv03}. Note that the secondary spectral component has a blue shift relative to the systemic velocity.
We show the orbital solutions for j06 and j09 in Figures~\ref{fig:lcrv06} and \ref{fig:lcrv09}. The agreement with a circular orbit is good, while $\rv$ data for the secondary spectral component are almost constant, with a slight red shift relative to systemic velocity. The radial velocities from \cite{kounkel} also agree with with our solution. For j09 there is a small signature of the Rossiter-McLaughlin (RM) effect \citep[][]{rossiter,mclaflin} around phase 1.0, although the $\rv$ uncertainties are too large to tell if it is real or not. High-resolution spectral observations can be very useful for this. Also the LC doesn't show any eclipse in this system, so the RM effect is very unlikely to be seen.

\begin{table*}
    \centering
    \begin{tabular}{c|ccc}
\hline
\hline
star & j03 & j06 & j09\\
\hline 
    $N_\rv$ & 13 & 60 & 63\\
    $P$, d & $5.730917 \pm   0.001028$ & $0.477201 \pm 0.000002$& $1.310644 \pm 0.000003$\\
    $t_0$, HJD d & $2458806.301979\pm0.037755$ & $2458496.584909\pm0.001147$& $2458468.635715\pm0.000843$\\

    $\gamma,\,\kms$ & $14.81\pm1.79$ & $22.95\pm0.72$& $32.69\pm0.24$\\
    $K,\,\kms$ & $61.15\pm2.53$ & $68.60\pm1.02$& $84.82\pm0.34$\\
\hline
    \end{tabular}
    \caption{Orbital solutions with {\sc GLS}.}
    \label{tab:gls}
\end{table*}

\subsection{Light curves analysis}

\label{sec:lc}

\begin{figure}
    \centering
    \includegraphics[width=\columnwidth]{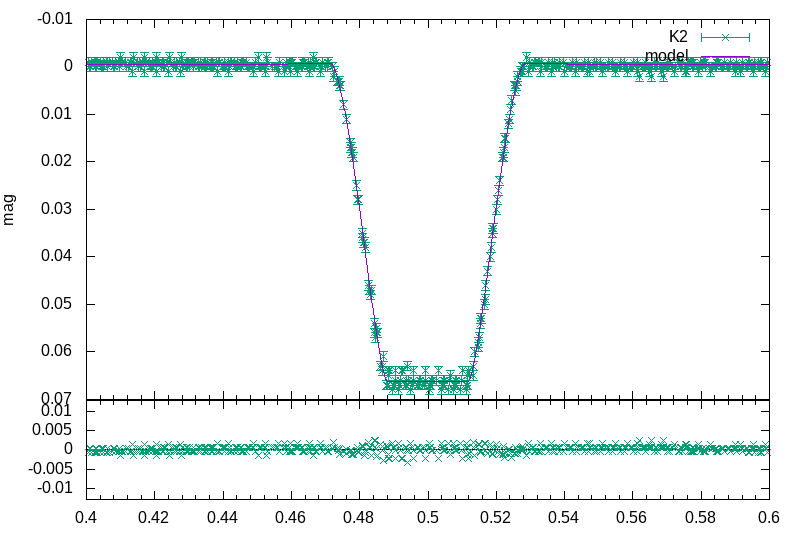}
    \includegraphics[width=\columnwidth]{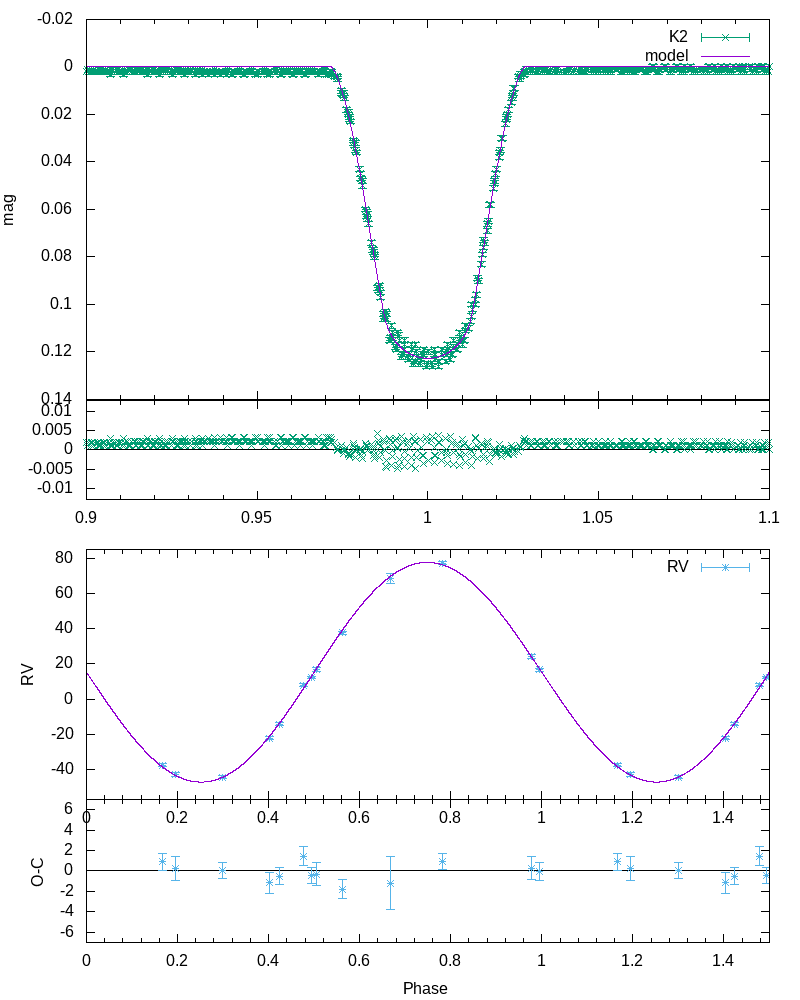}
    \caption{Best fit {\sc JKTEBOP} model for j03. Top and middle panels show eclipses, bottom panel shows $\rv_1$.}
    \label{fig:jkt}
\end{figure}

Among our three systems only LC for j03 shows eclipses (one of  them is total, when the smaller component is completely hidden), thus we decide to model it using similar methods to ones in \cite{j0647}. 
We used the \texttt{JKTEBOP} code (version 40)\footnote{{\sc jktebop} is written in {\sc fortran77} and the source code is available at \url{http://www.astro.keele.ac.uk/jkt/codes/jktebop.html}} by \citet{jkt} to simultaneously fit the LC and RV timeseries in mode ``task 3". Our fitting was initialised using $P,t_0,K,\gamma$ values from the \texttt{GLS} fit. We used limb darkening coefficients provided by the {\sc JKTLD} code and linearly interpolated them for spectral parameters of the primary: ${\teff}_A=6300$ K, $\logg=4.0$ dex, [M/H]=-0.3 dex. For the secondary component we set both coefficient to 0.3. We used quadratic limb darkening coefficients ($LD$). The gravity darkening coefficients were set to $\alpha=0.32$ for both components. The reflection coefficients were computed based on system's geometry.
The radial velocity was fitted for the primary component. Additionally we fit for a ``third" light contribution $L_3$, the nuisance parameter, the out-of-eclipse magnitude $S_0$. In total we fit for 12 parameters: $J$ the central surface brightness ratio, $(R_1+R_2)/a$ the ratio of the sum of stellar radii to the semimajor axis, $R_2/R_1$ the ratio of the radii, $i$ the inclination, $e\cos{\omega}$, $e\sin{\omega}$  the eccentricity multiplied by the cosine and sine of the periastron longitude, $P$ the period, $t_0$, semiamplitude and systemic velocity $K_{1}$, $\gamma_{1}$, $L_3$ and $S_0$. We use integration ring size $1^{\circ}$. The best fit parameters are listed in Table~\ref{tab:orbit} and Figure~\ref{fig:jkt} shows the fit.

\begin{table}

\begin{center}

\caption{{\sc JKTEBOP} solution for j03}

\begin{tabular}{lc}
\hline
Parameter & Value\\
\hline
fitted:\\
$J$  &0.636$\pm$0.003 \\
 $(R_1+R_2)/a$  &0.179$\pm$0.001 \\
 $R_2/R_1$  &0.412$\pm$0.009 \\
 $i^{\circ}$  &89.77$\pm$2.35 \\
 $e\cos{\omega}$  &-0.0001$\pm$0.0001 \\
 $e\sin{\omega}$  &-0.0050$\pm$0.0001 \\
 $P$,d  &5.73161$\pm$0.00001 \\
 $t_0$, HJD d& 2458806.25935$\pm$0.00353 \\
 $S_0$, mag & $0.0006 \pm  0.2515$\\
 $L_3$ & $0.399 \pm  0.024$\\
 $K_1,\,\kms$ &62.39$\pm$0.44 \\
 $\gamma_1,\,\kms$ &15.28$\pm$0.38 \\
 fixed\\
 ${\rm reflected~light}_1$, mag  &0.0006 \\
 ${\rm reflected~light}_2$, mag  &0.0010 \\
quadratic $LD_1$ & 0.2992, 0.3087\\
quadratic $LD_2$ & 0.3000, 0.3000\\
\hline
\label{tab:orbit}

\end{tabular}

\end{center}

\end{table}

\begin{figure}
    \centering
    \includegraphics[width=\columnwidth]{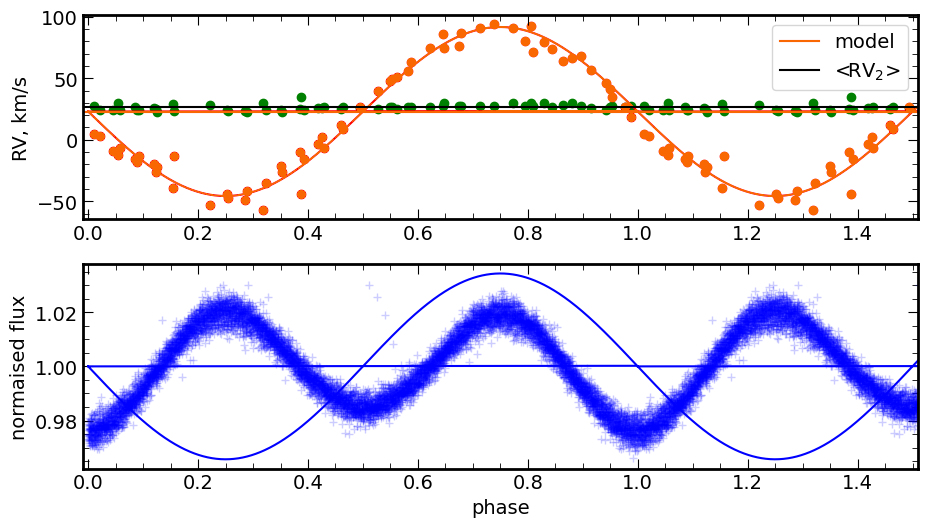}
    \caption{TESS LC for j06 folded with the period from orbital solution, which is schematically shown on the bottom panel. RV data (orange - $\rv_1$, green - $\rv_2$) with best-fit circular orbit (top panel) }
    \label{fig:lcrv06}
\end{figure}

\begin{figure}
    \centering
    \includegraphics[width=\columnwidth]{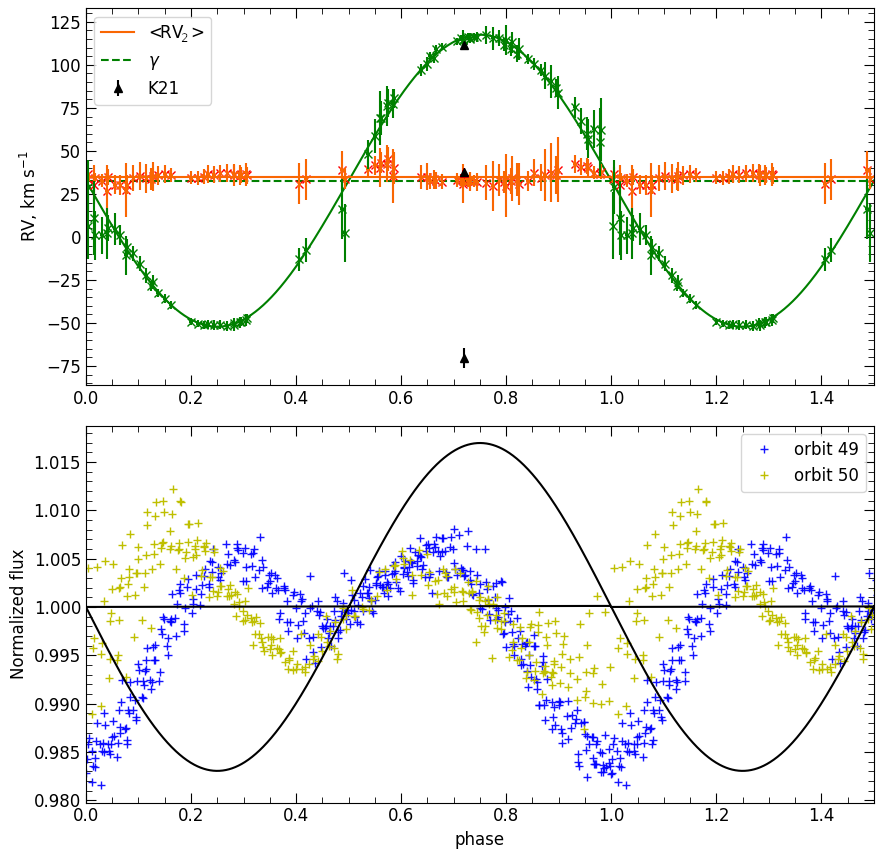}
    \caption{TESS LC for j09 folded with period from orbital solution, which is schematically shown on the bottom panel. RV data (green - $\rv_1$, orange - $\rv_2$) with best-fit circular orbit (top panel). RV data from \protect\cite{kounkel} are shown as black triangles.}
    \label{fig:lcrv09}
\end{figure}

The LCs for j06 and j09 are folded with periods from the orbital solution and shown in Figures~\ref{fig:lcrv06} and \ref{fig:lcrv09}. The photometrical data agree with the orbital solutions from the previous section. Changes in the LC are smaller than 3 per cent and there are no eclipses. The shape of the LCs suggests that the primary components in these two systems are non-spherical and significantly distorted by the tidal forces: the LC maxima correspond to $\rv_{\rm min,max}$ - moments when we can observe maximal projection of the primary component on the sky-plane. The LC minima have different depth: thus the primary's side facing the secondary is cooler due to gravitational darkening (highly likely for j06) or the secondary component reflects light from the primary (highly likely for j09). We leave detailed modelling of the LCs of these two systems for future studies.  
\par
Also we should note, that for j09 the data collected during TESS orbit 50 is very different from orbit 49, which can support the conclusion of \cite{asassn2023} that j09 is a spotted star. However it can be the same problem with data reduction (contamination and normalisation) as one mentioned in \cite{tyc}, where we decided to not use data from orbit 50 completely. 

\section{Discussion}
\label{discus}

Analysis of these three systems confirms that the Wilson method cannot be applied to their $\rv$ measurements as they are not simple SB2. Below we will discuss them in detail. 

\subsection{j03}

This system is a SB3 with significant difference in luminosity between components: the primary ($\sim55$ per cent) and secondary ($\sim5$ per cent) form an inner close binary subsystem with period $P\sim5.73$ d, while the third component ($\sim 40$ percent) has an almost constant $\rv$, with a slight blue shift ($\leq10~\kms$) relative to the systemic velocity of the inner system, see Figure~\ref{fig:rv03}, where it is shown as orange circles. This $\rv$ shift is small, so we think the third component is bound to the inner system. The simple two body orbital solution cannot capture the possible gravitational influence of third component, thus we recommend to use a more complex model in any future analysis of this system. If it is a triple system light-travel-time effect \citep{ltte} should also be observed. We hope that such analysis will be done in future studies of this system, when longer time series will be available.  
\par
Thanks to spectral disentangling results for $K_{1,2}$ we can roughly estimate mass ratio in the inner subsystem $Q=K_1/K_2\sim0.63$. These results also allow us to roughly estimate the orbital semimajor axis $a$ and the total mass $M_{\rm tot}=(M_1+M_2)$ using the Kepler's third law:

 \begin{align}
     a=(K_1 + K_2)\frac{P}{2 \pi \sin{i}}=0.085~{\rm [a.u.]}=18.32~[R_\odot],\\
     M_{\rm tot}=\frac{(K_1+K_2)^3}{ GM_{\odot} \sin^3{i}} \frac {P}{2 \pi}=2.51~[M_{\odot}] ,
     \label{eq:kepler3}
 \end{align}
where $GM_\odot=1.32712440041\cdot 10^{20}\, {\rm m^3\,s^{-2}}$  is the Solar mass parameter\footnote{\url{https://iau-a3.gitlab.io/NSFA/NSFA_cbe.html\#GMS2012}} and $i$ is the inclination of the orbit to the sky-plane. Masses of the components can be found using total mass and $Q$: $M_{1,2}=1.54,~0.97~M_\odot$. With the {\sc JKTEBOP} solution we also make a rough estimation of the stellar sizes: $R_{1,2}=2.32,~0.96~R_\odot$. Thus the secondary component is similar to the Sun, but it is completely out-shined by other two stars in the spectrum. All these estimates are rough as we don't have an error value for $K_2$ and ignore the possible influence of the third component. 


\subsection{j06}
The two visible spectral components belong to two stars separated by $2.5\arcsec$ observed by the same fiber of the LAMOST-MRS.
In Gaia DR2 \citep{gdr2} both components had very similar parallaxes $\varpi_{1,2}=1.1747\pm0.0476,~1.1744 \pm0.0449 $ mas, however recent Gaia DR3 values $\varpi_{1,2}=1.2137 \pm0.0175,~1.1445 \pm0.0248 $ mas suggest that the brighter component is a bit closer ($\sim50$ pc) to us, rejecting the possibility of them forming a wide system. We should note that $\langle\rv_2\rangle$ is very similar to the systemic velocity of the primary component (with red shift $\Delta\rv~\sim3.3~\kms$).  However proper motions from Gaia DR3 are quite different: $\mu_{\alpha} \cos{\delta}=0.674 \pm0.017,~1.472 \pm0.023~{\rm mas~yr^{-1}}$, $\mu_\delta=-3.924 \pm0.013,~-4.271 \pm0.018~{\rm mas~yr^{-1}}$ rejecting a possibility of j06 being a co-moving group of stars, which were born together. Thus we just have a chance alignment of SB1 and another star.
\par
Using the orbital solution, the binary mass function can be calculated as follows:
\begin{equation}
    f(M) = \frac{M_{2} \, \textrm{sin}^3 i} {(1+q)^{2}} = \frac{P \, K_{1}^{3} \, (1-e^2)^{3/2}}{2\pi GM_\odot}=0.0160\pm0.0007~[M_{\odot}] ,
\label{mass_function.equ}
\end{equation}
\noindent
where $M_{2}$ is the mass of the unseen star, $q = M_{1}/M_{2}$ is the mass ratio of this system, and $i$ is the orbital inclination. 
By using the mass $M_1=1.489^{+0.040}_{-0.042}~M_{\odot}$ of the visible star from Gaia DR3, we determined the minimum mass (taking $i=90^\circ$) of the unseen object as $0.36\pm0.02~M_{\odot}$. Therefore we think that the unseen secondary is possibly a white dwarf. 
\par
Clearly visible ellipsoidal variability in the LC tell us that the primary component is significantly distorted by tidal forces. Spots are unlikely to cause such variability as it is very stable over full observation interval (150 periods) of this system by TESS, see Figure~\ref{fig:lcrv06}. Clear agreement between phased RV and LC time series also supports that variability is coming from the system in question, not from the other nearby stars observed by TESS.  
\subsection{j09}

\begin{figure}
    \centering
    \includegraphics[width=\columnwidth]{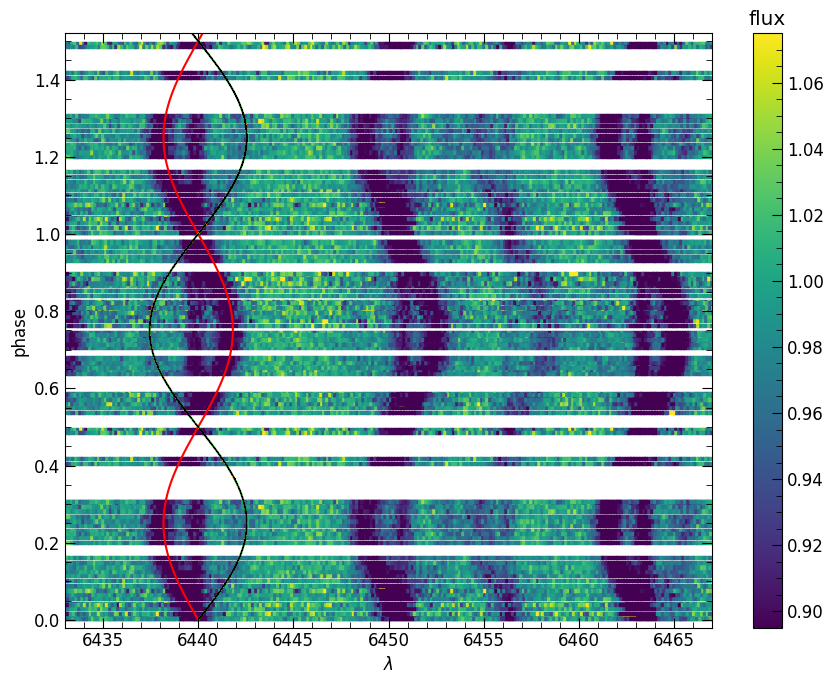}
    \includegraphics[width=\columnwidth]{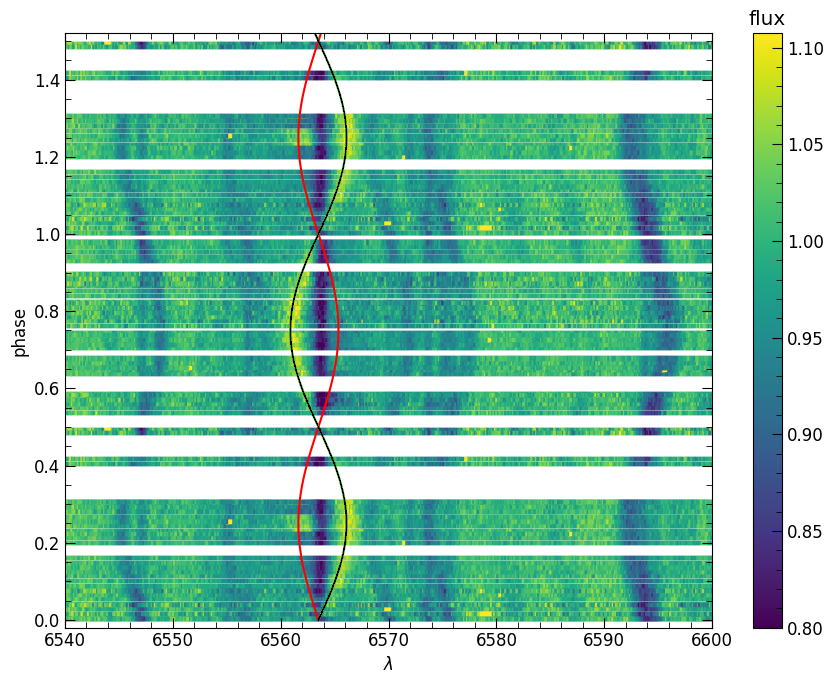}
    \caption{Dynamic spectrum of j09 in two regions of the red arm. The expected positions of the lines based on the orbital solution are shown with red (primary) and  black (secondary) lines.}
    \label{fig:red}
\end{figure}

This system is the only one, where we can see three components, see Figure~\ref{fig:spexampl}. Moreover, the secondary component has strong and stable emission in $\ha$ line, while this line is almost invisible in the primary, see the dynamic spectrum in the bottom panel of Figure~\ref{fig:red}.
Also note that at phase $\phi=0.25$ the primary has emission in $\ha$. Such variable emission can support chromospheric activity of this component, which was also seen as spots by \cite{asassn2023}.
Like in j06, the third spectral component has a slight red shift relative to the systematic velocity ($\Delta\rv~\sim2~\kms$), which can indicate that all three spectral components can form a large triple system. Unfortunately we didn't find any periodic changes in it's $\rv$ to confirm this hypothesis. 
\par 
Thanks to the spectral disentangling results for $K_{1,2}$ we can roughly estimate the mass ratio of the inner subsystem $Q=K_1/K_2\sim0.73$. These results also allow us to make a rough estimation of $a$ and $M_{\rm tot}$:

 \begin{align}
     a~\sin{i}=(K_1 + K_2)\frac{P}{2 \pi }=0.024~{\rm [a.u.]}=5.13~[R_\odot],\\
     M_{\rm tot} \sin^3{i}=\frac{(K_1+K_2)^3}{ GM_{\odot} } \frac {P}{2 \pi}=1.06~[M_{\odot}] ,
     \label{eq:kepler31}
\end{align}

Minimum masses of the components can be found using the total mass, $Q$ and $i=90^\circ$: $M_{1,2}=0.61,~0.45~M_\odot$. Thus they are highly likely red dwarfs.

\section{Conclusions}
\label{concl}

We selected three double-lined spectroscopic binary systems, which have extreme mass ratios, if measured using the Wilson method. The secondary spectral component in all of them has almost constant radial velocity. We analysed medium resolution spectroscopic observations from LAMOST-MRS and space-based photometry from Kepler and TESS and find out, that all three systems are not simple SB2s, but rather triple systems and a chance alignment of another star with j06 that have an unseen component, possibly white dwarf. Thus we conclude that suspicious, extreme mass ratios measured using the Wilson method for SB2 systems are incorrect and these systems should be treated using more complex models.


\section*{Acknowledgements}
We are grateful to the anonymous referee for a constructive report.
We thank Hans B{\"a}hr for his careful proof-reading of the manuscript.
We thank Sa{\v s}a Iliji{\'c} for help with {\sc FD3} code.
We thank Hans Ludwig for useful discussion about nature of j09 system.
MK is grateful to his parents, Yuri Kovalev and Yulia Kovaleva and to his aunt Elena Krivkina, for their full support in making this research possible. The work is supported by the Natural Science Foundation of China (Nos. 11733008, 12090040, 12090043, 11521303, 12125303, 12273057, 12288102).
Guoshoujing Telescope (the Large Sky Area Multi-Object Fiber Spectroscopic Telescope LAMOST) is a National Major Scientific Project built by the Chinese Academy of Sciences. Funding for the project has been provided by the National Development and Reform Commission. LAMOST is operated and managed by the National Astronomical Observatories, Chinese Academy of Sciences. The authors gratefully acknowledge the “PHOENIX Supercomputing Platform” jointly operated by the Binary Population Synthesis Group and the Stellar Astrophysics Group at Yunnan Observatories, Chinese Academy of Sciences. 
This research has made use of NASA’s Astrophysics Data System, the SIMBAD data base, and the VizieR catalogue access tool, operated at CDS, Strasbourg, France. It also made use of TOPCAT, an interactive graphical viewer and editor for tabular data \citep[][]{topcat}.  Funding for the Kepler and TESS mission is provided by NASA’s Science Mission directorate. This paper includes data collected by the Kepler and TESS missions, which are publicly available from the Mikulski Archive for Space Telescopes (MAST). This work has made use of data from the European Space Agency (ESA) mission {\it Gaia} (\url{https://www.cosmos.esa.int/gaia}), processed by the {\it Gaia} Data Processing and Analysis Consortium (DPAC, \url{https://www.cosmos.esa.int/web/gaia/dpac/consortium}). Funding for the DPAC has been provided by national institutions, in particular the institutions participating in the {\it Gaia} Multilateral Agreement.

\section*{Data Availability}
The data underlying this article will be shared on reasonable request to the corresponding author.




\bibliographystyle{mnras}

\begin{thebibliography}{}
\makeatletter
\relax
\def\mn@urlcharsother{\let\do\@makeother \do\$\do\&\do\#\do\^\do\_\do\%\do\~}
\def\mn@doi{\begingroup\mn@urlcharsother \@ifnextchar [ {\mn@doi@}
  {\mn@doi@[]}}
\def\mn@doi@[#1]#2{\def\@tempa{#1}\ifx\@tempa\@empty \href
  {http://dx.doi.org/#2} {doi:#2}\else \href {http://dx.doi.org/#2} {#1}\fi
  \endgroup}
\def\mn@eprint#1#2{\mn@eprint@#1:#2::\@nil}
\def\mn@eprint@arXiv#1{\href {http://arxiv.org/abs/#1} {{\tt arXiv:#1}}}
\def\mn@eprint@dblp#1{\href {http://dblp.uni-trier.de/rec/bibtex/#1.xml}
  {dblp:#1}}
\def\mn@eprint@#1:#2:#3:#4\@nil{\def\@tempa {#1}\def\@tempb {#2}\def\@tempc
  {#3}\ifx \@tempc \@empty \let \@tempc \@tempb \let \@tempb \@tempa \fi \ifx
  \@tempb \@empty \def\@tempb {arXiv}\fi \@ifundefined
  {mn@eprint@\@tempb}{\@tempb:\@tempc}{\expandafter \expandafter \csname
  mn@eprint@\@tempb\endcsname \expandafter{\@tempc}}}

\bibitem[\protect\citeauthoryear{{Ahumada} et~al.,}{{Ahumada}
  et~al.}{2020}]{apogee16}
{Ahumada} R.,  et~al., 2020, \mn@doi [\apjs] {10.3847/1538-4365/ab929e}, \href
  {https://ui.adsabs.harvard.edu/abs/2020ApJS..249....3A} {249, 3}

\bibitem[\protect\citeauthoryear{{Barros}, {Demangeon}  \& {Deleuil}}{{Barros}
  et~al.}{2016}]{polarK2}
{Barros} S.~C.~C.,  {Demangeon} O.,   {Deleuil} M.,  2016, \mn@doi [\aap]
  {10.1051/0004-6361/201628902}, \href
  {https://ui.adsabs.harvard.edu/abs/2016A&A...594A.100B} {594, A100}

\bibitem[\protect\citeauthoryear{{Brasseur}, {Phillip}, {Fleming}, {Mullally}
  \& {White}}{{Brasseur} et~al.}{2019}]{astrocut}
{Brasseur} C.~E.,  {Phillip} C.,  {Fleming} S.~W.,  {Mullally} S.~E.,   {White}
  R.~L.,  2019, {Astrocut: Tools for creating cutouts of TESS images},
  Astrophysics Source Code Library, record ascl:1905.007 (\mn@eprint {ascl}
  {1905.007})

\bibitem[\protect\citeauthoryear{{Christy} et~al.,}{{Christy}
  et~al.}{2023}]{asassn2023}
{Christy} C.~T.,  et~al., 2023, \mn@doi [\mnras] {10.1093/mnras/stac3801},
  \href {https://ui.adsabs.harvard.edu/abs/2023MNRAS.519.5271C} {519, 5271}

\bibitem[\protect\citeauthoryear{{Cui} et~al.,}{{Cui}
  et~al.}{2012}]{2012RAA....12.1197C}
{Cui} X.-Q.,  et~al., 2012, \mn@doi [Research in Astronomy and Astrophysics]
  {10.1088/1674-4527/12/9/003}, \href
  {https://ui.adsabs.harvard.edu/abs/2012RAA....12.1197C} {12, 1197}

\bibitem[\protect\citeauthoryear{{Czesla}, {Schr{\"o}ter}, {Schneider},
  {Huber}, {Pfeifer}, {Andreasen}  \& {Zechmeister}}{{Czesla}
  et~al.}{2019}]{pya}
{Czesla} S.,  {Schr{\"o}ter} S.,  {Schneider} C.~P.,  {Huber} K.~F.,  {Pfeifer}
  F.,  {Andreasen} D.~T.,   {Zechmeister} M.,  2019, {PyA: Python
  astronomy-related packages} (\mn@eprint {ascl} {1906.010})

\bibitem[\protect\citeauthoryear{{Feinstein} et~al.,}{{Feinstein}
  et~al.}{2019}]{eleanor}
{Feinstein} A.~D.,  et~al., 2019, \mn@doi [\pasp] {10.1088/1538-3873/ab291c},
  \href {https://ui.adsabs.harvard.edu/abs/2019PASP..131i4502F} {131, 094502}

\bibitem[\protect\citeauthoryear{{Gaia Collaboration} et~al.,}{{Gaia
  Collaboration} et~al.}{2018}]{gdr2}
{Gaia Collaboration} et~al., 2018, \mn@doi [\aap]
  {10.1051/0004-6361/201833051}, \href
  {http://adsabs.harvard.edu/abs/2018A%26A...616A...1G} {616, A1}

\bibitem[\protect\citeauthoryear{{Gaia Collaboration} et~al.,}{{Gaia
  Collaboration} et~al.}{2022}]{gaia3}
{Gaia Collaboration} et~al., 2022, arXiv e-prints, \href
  {https://ui.adsabs.harvard.edu/abs/2022arXiv220800211G} {p. arXiv:2208.00211}

\bibitem[\protect\citeauthoryear{{Green}, {Maoz}, {Mazeh}, {Faigler}, {Shahaf},
  {Gomel}, {El-Badry}  \& {Rix}}{{Green} et~al.}{2023}]{green2023}
{Green} M.~J.,  {Maoz} D.,  {Mazeh} T.,  {Faigler} S.,  {Shahaf} S.,  {Gomel}
  R.,  {El-Badry} K.,   {Rix} H.-W.,  2023, \mn@doi [\mnras]
  {10.1093/mnras/stad915}, \href
  {https://ui.adsabs.harvard.edu/abs/2023MNRAS.522...29G} {522, 29}

\bibitem[\protect\citeauthoryear{{Grupp}}{{Grupp}}{2004a}]{Grupp2004a}
{Grupp} F.,  2004a, \mn@doi [\aap] {10.1051/0004-6361:20040971}, 420, 289

\bibitem[\protect\citeauthoryear{{Grupp}}{{Grupp}}{2004b}]{Grupp2004b}
{Grupp} F.,  2004b, \mn@doi [\aap] {10.1051/0004-6361:20040456}, 426, 309

\bibitem[\protect\citeauthoryear{{Hajdu}, {Mat{\'e}csa}, {Sallai}  \&
  {B{\'o}di}}{{Hajdu} et~al.}{2022}]{ltte}
{Hajdu} T.,  {Mat{\'e}csa} B.,  {Sallai} J.~M.,   {B{\'o}di} A.,  2022, \mn@doi
  [\mnras] {10.1093/mnras/stac2533}, \href
  {https://ui.adsabs.harvard.edu/abs/2022MNRAS.tmp.2394H} {}

\bibitem[\protect\citeauthoryear{{Huang} et~al.,}{{Huang}
  et~al.}{2020a}]{tess1}
{Huang} C.~X.,  et~al., 2020a, \mn@doi [Research Notes of the American
  Astronomical Society] {10.3847/2515-5172/abca2e}, \href
  {https://ui.adsabs.harvard.edu/abs/2020RNAAS...4..204H} {4, 204}

\bibitem[\protect\citeauthoryear{{Huang} et~al.,}{{Huang}
  et~al.}{2020b}]{tess2}
{Huang} C.~X.,  et~al., 2020b, \mn@doi [Research Notes of the American
  Astronomical Society] {10.3847/2515-5172/abca2d}, \href
  {https://ui.adsabs.harvard.edu/abs/2020RNAAS...4..206H} {4, 206}

\bibitem[\protect\citeauthoryear{{Iliji{\'c}}}{{Iliji{\'c}}}{2017}]{fd3}
{Iliji{\'c}} S.,  2017, {fd3: Spectral disentangling of double-lined
  spectroscopic binary stars}, Astrophysics Source Code Library, record
  ascl:1705.012 (\mn@eprint {ascl} {1705.012})

\bibitem[\protect\citeauthoryear{{Ilijic}, {Hensberge}, {Pavlovski}  \&
  {Freyhammer}}{{Ilijic} et~al.}{2004}]{fd3norm}
{Ilijic} S.,  {Hensberge} H.,  {Pavlovski} K.,   {Freyhammer} L.~M.,  2004, in
  {Hilditch} R.~W.,  {Hensberge} H.,   {Pavlovski} K.,  eds,  Astronomical
  Society of the Pacific Conference Series Vol. 318, Spectroscopically and
  Spatially Resolving the Components of the Close Binary Stars. pp 111--113

\bibitem[\protect\citeauthoryear{{Kounkel} et~al.,}{{Kounkel}
  et~al.}{2021}]{kounkel}
{Kounkel} M.,  et~al., 2021, \mn@doi [\aj] {10.3847/1538-3881/ac1798}, \href
  {https://ui.adsabs.harvard.edu/abs/2021AJ....162..184K} {162, 184}

\bibitem[\protect\citeauthoryear{{Kovalev}}{{Kovalev}}{2019}]{disser}
{Kovalev} M.,  2019, PhD thesis, IMPRS-HD, \mn@doi{10.11588/heidok.00027411}

\bibitem[\protect\citeauthoryear{{Kovalev}, {Li}, {Zhang}, {Li}, {Chen}  \&
  {Han}}{{Kovalev} et~al.}{2022a}]{tyc}
{Kovalev} M.,  {Li} Z.,  {Zhang} X.,  {Li} J.,  {Chen} X.,   {Han} Z.,  2022a,
  \mn@doi [\mnras] {10.1093/mnras/stac1177}, \href
  {https://ui.adsabs.harvard.edu/abs/2022MNRAS.513.4295K} {513, 4295}

\bibitem[\protect\citeauthoryear{{Kovalev}, {Chen}  \& {Han}}{{Kovalev}
  et~al.}{2022b}]{cat22}
{Kovalev} M.,  {Chen} X.,   {Han} Z.,  2022b, \mn@doi [\mnras]
  {10.1093/mnras/stac2513}, \href
  {https://ui.adsabs.harvard.edu/abs/2022MNRAS.517..356K} {517, 356}

\bibitem[\protect\citeauthoryear{{Kovalev}, {Wang}, {Chen}  \& {Han}}{{Kovalev}
  et~al.}{2023}]{j0647}
{Kovalev} M.,  {Wang} S.,  {Chen} X.,   {Han} Z.,  2023, \mn@doi [\mnras]
  {10.1093/mnras/stac3767}, \href
  {https://ui.adsabs.harvard.edu/abs/2023MNRAS.519.5454K} {519, 5454}

\bibitem[\protect\citeauthoryear{{Liu} et~al.,}{{Liu} et~al.}{2020}]{mrs}
{Liu} C.,  et~al., 2020, arXiv e-prints, \href
  {https://ui.adsabs.harvard.edu/abs/2020arXiv200507210L} {p. arXiv:2005.07210}

\bibitem[\protect\citeauthoryear{{McLaughlin}}{{McLaughlin}}{1924}]{mclaflin}
{McLaughlin} D.~B.,  1924, \mn@doi [\apj] {10.1086/142826}, \href
  {https://ui.adsabs.harvard.edu/abs/1924ApJ....60...22M} {60, 22}

\bibitem[\protect\citeauthoryear{{Olivares} et~al.,}{{Olivares}
  et~al.}{2018}]{j03membership}
{Olivares} J.,  et~al., 2018, \mn@doi [\aap] {10.1051/0004-6361/201730972},
  \href {https://ui.adsabs.harvard.edu/abs/2018A&A...617A..15O} {617, A15}

\bibitem[\protect\citeauthoryear{{Ricker} et~al.,}{{Ricker}
  et~al.}{2015}]{tess}
{Ricker} G.~R.,  et~al., 2015, \mn@doi [Journal of Astronomical Telescopes,
  Instruments, and Systems] {10.1117/1.JATIS.1.1.014003}, \href
  {https://ui.adsabs.harvard.edu/abs/2015JATIS...1a4003R} {1, 014003}

\bibitem[\protect\citeauthoryear{{Rossiter}}{{Rossiter}}{1924}]{rossiter}
{Rossiter} R.~A.,  1924, \mn@doi [\apj] {10.1086/142825}, \href
  {https://ui.adsabs.harvard.edu/abs/1924ApJ....60...15R} {60, 15}

\bibitem[\protect\citeauthoryear{{Southworth}}{{Southworth}}{2013}]{jkt}
{Southworth} J.,  2013, \mn@doi [\aap] {10.1051/0004-6361/201322195}, \href
  {https://ui.adsabs.harvard.edu/abs/2013A&A...557A.119S} {557, A119}

\bibitem[\protect\citeauthoryear{{Stassun} et~al.,}{{Stassun}
  et~al.}{2019}]{tic}
{Stassun} K.~G.,  et~al., 2019, \mn@doi [\aj] {10.3847/1538-3881/ab3467}, \href
  {https://ui.adsabs.harvard.edu/abs/2019AJ....158..138S} {158, 138}

\bibitem[\protect\citeauthoryear{{Taylor}}{{Taylor}}{2005}]{topcat}
{Taylor} M.~B.,  2005, in {Shopbell} P.,  {Britton} M.,   {Ebert} R.,  eds,
  Astronomical Society of the Pacific Conference Series Vol. 347, Astronomical
  Data Analysis Software and Systems XIV. p.~29

\bibitem[\protect\citeauthoryear{{Ting}, {Conroy}, {Rix}  \& {Cargile}}{{Ting}
  et~al.}{2019}]{ting2019}
{Ting} Y.-S.,  {Conroy} C.,  {Rix} H.-W.,   {Cargile} P.,  2019, \mn@doi [\apj]
  {10.3847/1538-4357/ab2331}, \href
  {https://ui.adsabs.harvard.edu/abs/2019ApJ...879...69T} {879, 69}

\bibitem[\protect\citeauthoryear{{Watson}, {Henden}  \& {Price}}{{Watson}
  et~al.}{2006}]{varstarindex}
{Watson} C.~L.,  {Henden} A.~A.,   {Price} A.,  2006, Society for Astronomical
  Sciences Annual Symposium, \href
  {https://ui.adsabs.harvard.edu/abs/2006SASS...25...47W} {25, 47}

\bibitem[\protect\citeauthoryear{{Wilson}}{{Wilson}}{1941}]{wilson}
{Wilson} O.~C.,  1941, \mn@doi [\apj] {10.1086/144239}, \href
  {https://ui.adsabs.harvard.edu/abs/1941ApJ....93...29W} {93, 29}

\bibitem[\protect\citeauthoryear{{Zechmeister} \& {K{\"u}rster}}{{Zechmeister}
  \& {K{\"u}rster}}{2009}]{gls}
{Zechmeister} M.,  {K{\"u}rster} M.,  2009, \mn@doi [\aap]
  {10.1051/0004-6361:200811296}, \href
  {https://ui.adsabs.harvard.edu/abs/2009A&A...496..577Z} {496, 577}

\bibitem[\protect\citeauthoryear{{Zhao}, {Zhao}, {Chu}, {Jing}  \&
  {Deng}}{{Zhao} et~al.}{2012}]{2012RAA....12..723Z}
{Zhao} G.,  {Zhao} Y.-H.,  {Chu} Y.-Q.,  {Jing} Y.-P.,   {Deng} L.-C.,  2012,
  \mn@doi [Research in Astronomy and Astrophysics]
  {10.1088/1674-4527/12/7/002}, \href
  {https://ui.adsabs.harvard.edu/abs/2012RAA....12..723Z} {12, 723}

\makeatother
\end{thebibliography}





\appendix

\section{Spectral models}
\label{sec:payne}
The synthetic spectra are generated using NLTE~MPIA online-interface \url{https://nlte.mpia.de} \citep[see Chapter~4 in][]{disser} on wavelength intervals 4870:5430 \AA~for the blue arm and 6200:6900 \AA ~for the red arm with spectral resolution $R=7500$. We use NLTE (non-local thermodynamic equilibrium) spectral synthesis for H, Mg~I, Si~I, Ca~I, Ti~I, Fe~I and Fe~II lines \citep[see Chapter~4 in][ for references]{disser}.  
\par
The grid of models (6200 in total) is computed for points randomly selected in a range of $\teff$ between 4600 and 8800 K, $\logg$ between 1.0 and 4.8  (cgs units), $\vsini$ from 1 to 300 $\kms$ and [Fe/H]\footnote{We used $\feh$ as a proxy of overall metallicity, abundances for all elements are scaled with Fe.} between $-$0.9 and $+$0.9 dex. The model is computed only if linear interpolation of MAFAGS-OS\citep[][]{Grupp2004a,Grupp2004b} stellar atmosphere is possible for a given point in parameter space.  Microturbulence is fixed to $\Vmic=2~\kms$ for all models. The grid is randomly split on training (70\%) and cross-validation (30\%) sets of spectra, which are used to train \textit{The~Payne} spectral model \citep{ting2019}. The neural network (NN) consists of two layers of 300 neurons each with rectilinear unit (ReLU)\footnote{ReLU(x)=max(x,0)} activation functions. We train separate NNs for each spectral arm. The median approximation error is less than 1\% for both arms. We use output of \textit{The Payne} as single-star spectral model.   

\section{RV measurements}
 We provide the RV measurements used to fit orbital parameters in Table~\ref{tab:rvs}.
\begin{table}
    \centering
    \caption{\label{tab:rvs} Radial velocity measurements for primary components. Full table is available online}
    \begin{tabular}{lcc}
\hline
time HJD &  value & error  \\
 d    & $\kms$ & $\kms$\\

\hline
j03\\
\hline
2458802.242 &-44.304 &0.780 \\
2458820.152 &-14.188 &0.817 \\
2458830.141 &-37.588 &0.846 \\
2458836.033 &-42.888 &1.171 \\
2458852.089 &16.806 &0.880 \\
2458883.994 &37.598 &0.965 \\
2458890.986 &77.144 &0.780 \\
2459124.326 &12.263 &0.784 \\
2459148.249 &68.523 &2.610 \\
2459181.120 &-22.078 &1.033 \\
2459190.151 &24.035 &1.120 \\
2459216.095 &16.747 &1.126 \\
2459531.178 &7.846 &0.909 \\
.. & .. & ..\\
 \hline
\end{tabular}
\end{table}


\bsp	
\label{lastpage}
\end{document}